# Arbitrarily Accurate Classification Applied to Specific Emitter Identification

Michael C. Kleder

**Abstract**— This article introduces a method of evaluating subsamples until any prescribed level of classification accuracy is attained, thus obtaining arbitrary accuracy. A logarithmic reduction in error rate is obtained with a linear increase in sample count. The technique is applied to specific emitter identification on a published dataset of physically recorded over-the-air signals from 16 ostensibly identical high-performance radios. The technique uses a multi-channel deep learning convolutional neural network acting on the bispectra of I/Q signal subsamples each consisting of 56 parts per million (ppm) of the original signal duration. High levels of accuracy are obtained with minimal computation time: in this application, each addition of eight samples decreases error by one order of magnitude.

—————————— ◆ ——————————

## 1 INTRODUCTION

Specific emitter identification (SEI) is the process of uniquely labeling an individual radio as the source of a received signal by recognizing unintentionally emitted signal modulation, sometimes called an "RF fingerprint." This fingerprint arises from manufacturing deviations in the radio electronics, and allows one to distinguish it from other emitters which may be of the exact same design and model. Successful SEI allows military and commercial stakeholders to identify particular emitter devices even if they change locations in between observations, and thus to correlate emitter-specific information, such as digital signatures, encryption types, or hosting platforms, to observed signals.

In this paper, we investigate a published data set [1] of over-the-air signals from 16 identical USRP X310 software-defined radios which transmitted identical information using the same IEEE 802.11a modulation scheme. These radios are described by the manufacturer Ettus Research as "high performance" and have a retail price of about $7,000 each [2]. The data set was recorded by the Genesys research lab at Northeastern University and made available online in 2019 as the "Oracle RF Fingerprinting Dataset" [3]. In research funded by the Defense Advanced Research Projects Agency (DARPA) under its Radio Frequency Machine Learning Systems [4] program, the providers of the data set created a deep-learning system to identify specific radios with an error rate of 1.4% without interacting with them, in an accompanying paper [5]. They also obtained improved error rates of 0.5% and 0.24% when interacting with the radios and when introducing additional hardware impairments to them.

## 2 I/Q DATA

Radio frequency signals are created to occupy a bandwidth between some lower frequency and some higher frequency. An informational signal is transmitted by using it to change, or modulate, one or more "carrier" frequencies within that bandwidth using one of many available modulation schemes. One way of recovering the modulating information is to select a center frequency within the bandwidth and mathematically extracting the modulating signal relative to that center frequency. This is convenient because one then only needs to deal with the low-frequency modulating information, rather than the high-frequency carrier. The resulting signal has a new center frequency of zero, and thus distinct positive and negative frequency components. Because the Fourier transform of a real-valued signal must have symmetric positive and negative frequency components, a signal which has different positive and negative frequency components must be a complex-valued signal, and therefore can conveniently be expressed using real and imaginary component signals. The conventional way of expressing these two components describes the real and imaginary components as "in-phase" and "quadrature" signals, or I/Q data. With a center frequency of $f_0$, the in-phase and quadrature components of a signal $x(t)$ are obtained by [6]:

$$I(t) = x(t)\cos 2\pi f_0 t + H[x(t)]\sin 2\pi f_0 t$$

$$Q(t) = H[x(t)]\cos 2\pi f_0 t - x(t)\sin 2\pi f_0 t$$

where the Hilbert transform $H[x(t)]$ of $x(t)$ is defined [7] as:

$$H[x(t)] = \frac{1}{\pi}\int_{-\infty}^{\infty}\frac{x(\tau)}{t-\tau}d\tau = \mathcal{F}^{-1}\big[(-i\cdot\text{sign}(f))\cdot\mathcal{F}[x](f)\big]$$

———————————————
- *Michael C. Kleder is a doctoral candidate at the Grado Department of Industrial and Systems Engineering of Virginia Polytechnic Institute and State University, Blacksburg, VA 24061. Email: mkleder@gmail.com*

where $\mathcal{F}$ is the Fourier transform, and the integral is taken in Cauchy's principal value sense, meaning that the singularity where t = τ is ignored.

A significant benefit of conducting SEI analysis on I/Q data is that the analysis is independent of the frequency band of the signal, utilizing instead only its modulation information. This implies that an emitter cannot hide its identity by switching frequency bands, or alternatively that emitter modes can be more readily distinguished by utilizing the ability, conferred by the separation of I/Q and $f_0$, to consider the modulation and the frequency band as independent characteristics.

## 3 BISPECTRA

Unintentional modulation is created when manufacturing variances or production defects in electronics alter an emitted signal through the introduction of generally non-linear artifacts. This can be observed as a pattern of correlations in the appearance of pairs of frequencies within the width of the band in which a transmitter operates. Such a pattern is called a bispectrum. Unintentional modulation is characteristic of the physical makeup of an emitter, so it is generally consistent throughout an emitted signal regardless of different data that the emitter might be transmitting at various times. The signed amplitude of the bispectrum of a discretely sampled complex signal x(t) at angular frequencies $\omega_1$ and $\omega_2$ is defined [8] as the Fourier transform of the third-order cumulant of x(t) as:

$$B(\omega_1, \omega_2) = \sum_{\tau_1}\sum_{\tau_2} e^{-j(\omega_1\tau_1 + \omega_2\tau_2)} E_t[x^*(t)x(t+\tau_1)x(t+\tau_2)]$$

Because the cumulant is third-order (detecting skew), the bispectrum inherently ignores Gaussian background noise (the distribution of which has no skew) and de-emphasizes intentional modulation to the extent that the distribution of encoded data is symmetric (e.g., a sequence of informative binary digits would normally be expected to have a nearly symmetric distribution of zeros and ones). If the signal-generating system were entirely linear – that is, the signal is generated pursuant to a linear differential equation – then the solution signal, neglecting the possibility of non-physical exponential growth and decay terms, would consist entirely of sinusoidal components, and again the distribution of signed signal amplitudes would be a symmetric distribution of positive and negative values. This leaves the bispectrum with the skewed distribution of non-linear attributes of the signal generating system, and hence emphasizes the unintentional modulation arising from imperfections in the manufactured components that are characteristic of a specific emitter.

In a contrasting application such as computer vision, the automated system is often compared to the performance of a human being. For the ImageNet [9] database of images in 1,000 classes, the best accuracy obtained by a human who is permitted five guesses for each image is over 99%, while a typical accuracy currently obtained by an artificial intelligence (AI) is about 97% [10], and on those occasions when some AI outperforms some human, it is dubbed "superhuman." In contrast, adequate human performance is not expected to be possible for classification of bispectra, which are entirely abstract from the perspective of human perception, as shown by the notional bispectrum magnitude plot shown below. Therefore, AI classification of bispectra with any useful performance level will be "superhuman."

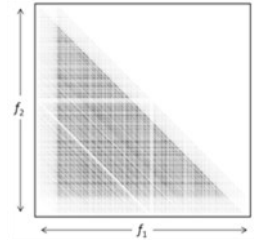

## 3 DEEP LEARNING

One of the tools of machine learning (ML) within the broader field of AI is the artificial neural network, which is a collection of interconnected mathematical primitives called neurons, which work together to model complex transformations from system inputs to system outputs. One type of neural network is the convolutional neural network, which scans a viewing aperture, typically across an image, and draws conclusions from basic image structures ("image primitives") which are detected. Subsequent convolutions scan the relative locations of these primitives, and then the relative locations of those aggregates, until an understanding of the entire image is attained. Neural networks are trained by adjusting the internal parameters of each neuron in response to whether the final conclusion of the overall network is correct or incorrect. We refer to a neural network as a "deep learning" network when there are many layers – typically more than five or six – which consecutively process the conclusions of prior layers. A modern deep learning network has hundreds of layers.

## 3 BINARY CLASSIFICATION BY VOTING

The "wisdom of crowds" can be simplistically interpreted as nomenclature for the fact that measurement error is typically Gaussian, and the mean of a large number of measurements is likely to be closer to the truth than any one particular measurement. If the measurement in question is a classification decision of one particular object into one of two categories, and if the classification decision is modeled as random, with the probability of classification of that one particular object into the first of the two categories provided by a corresponding fixed probability number $p$, then the distribution of $n$ classification decisions, each made by one member of the crowd of $n$ evaluators, follows a binomial distribution, where $k$ is the number of votes for the first class whose probability was $p$.

$$f_{binomial}(k, n, p) = \frac{n!}{k!\,(n-k)!} \cdot p^k(1-p)^{n-k}$$

As the size of the crowd increases, it becomes increasingly certain that the category containing the largest number of individual classification "votes" will be that category whose probability is the higher. If we suppose that we have as known a large set of votes, but we do not know what the underlying probability is, then we can think of the probability itself as an unknown (random, to us) entity, and its distribution is known as a Beta distribution. The density $f$ of probability $p$ for the first category, after ($\alpha$-1) votes for the first category and ($\beta$-1) votes for the second category, is as follows. (The use of ($\alpha$-1) and ($\beta$-1) is by convention, and the exclamation point represents the factorial operation.)

$$f_{beta}(p;\alpha,\beta) = \frac{(\alpha+\beta-1)!}{(\alpha-1)!\cdot(\beta-1)!} \cdot p^{(\alpha-1)}(1-p)^{(\beta-1)}$$

We observe that this is a probability distribution for the probability of the first category. Conceptually, the probability of the first category is a fixed quantity, but we simply don't know what it is, so we are modeling what we know about it. The benefit of thinking of the probability of the first category as following a distribution, rather than simply tabulating the largest number of votes and assessing a "winner," is that the Beta distribution also informs us of how certain we can be about the identity of the "winner," which allows us to continue counting votes until the certainty reaches some pre-determined threshold. We would not know how certain we can be about the identity of the "winning" category without this theoretical help. For example, suppose we want to be 99% certain that the winner of our voting scheme, is indeed the most probable category. If we observe only a single vote, we can declare a winner but we will not be able to assert anything about our certainty that the winning category had indeed been the most probable. However, according to the Beta distribution, if we have 60 votes for the first category and 40 votes for the second category, then $\alpha$ = 61 and $\beta$ = 41, giving

$$f_{beta}(p;61,41) = \frac{(61+41-1)!}{(61-1)!\cdot(41-1)!} \cdot p^{(61-1)}(1-p)^{(41-1)}$$

which we plot as follows for density f as a function of probability p.

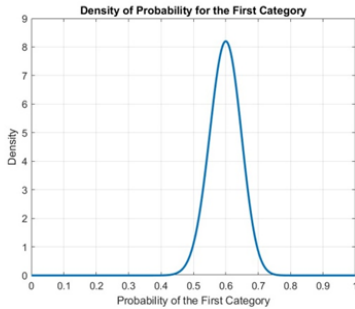

The question is, with 100 votes tallied, how certain are we that the first category is the "true" winner among a large population, and not just the winner of our 100-person sample, by random chance in the selection of the sample? Stated differently, what is the chance that the first category has a "true" probability of less than 50%? The cumulative Beta distribution F(p;$\alpha$,$\beta$) is:

$$F_{beta}(p;\alpha,\beta) = \frac{(\alpha+\beta-1)!}{(\alpha-1)!\cdot(\beta-1)!} \cdot \int_0^p x^{(\alpha-1)}(1-x)^{(\beta-1)}\,dx$$

In our example, $F(p;\alpha,\beta)$ = F(.5;61,41) = 2.3%, yielding a 97.7% certainty that the first category is truly the most favored.

A key insight at this point is that we can keep on tabulating more and more votes until our sample indicates the true winner with any degree of certainty we desire. In our example, if we quadruple the number of tabulated votes from 100 to 400, and still observed the same 60%/40% split in votes, we would be 99.997% certain that the first category would truly be more likely, because $F(.5;241,161)$ = 0.00003067.

The use of the cumulative Beta distribution on each category individually is a conservative approach, because it works even if the probability of the most probable category only minimally exceeds 50% for the most favored category, and the second most probable category consumes all of the remaining probability. In practice, the most probable category in our system attains, on average, a probability exceeding 80% (often significantly higher), and the second most probable category attains, on average, a probability for each single voter that does not exceed 16% (often significantly lower), because the probability for an incorrect category is divided among the several incorrect categories. As a result, when we choose a maximum acceptable error rate (such as 5%) we observe in practice a realized error rate that is much smaller (such as $10^{-5}$). We do not rely on this benefit, however. We utilize the conservative individual cumulative Beta function alone, to ensure that the error rate is below the specified threshold even in the worst cases.

## 4 MULTIPLE CATEGORY CLASSIFICATION BY VOTING

If the measurement in question is a classification decision of one particular object into one of several (N) categories, and if the classification decision is modeled as random, with the probability of classification of that one particular object into each category provided by a corresponding entry in a fixed vector of probability values, then the distribution of classification decisions made by a crowd of evaluators follows what is termed a multinomial distribution. Here, the vector $p$ contains and entry $p_i$ for each category $i$ in the set of categories {1, … , N}, and the number in that entry is the probability of that category being selected. The vector $k$ also contains an entry $k_i$ for each category $i$, and the number in that entry is the number of classification decisions that were made into that category out of n total decisions by the sample of voters. The multinomial distribution $f(k,n,p)$ is then

$$f_{multinomial}(\boldsymbol{k},n,\boldsymbol{p}) = \frac{n!}{\prod_{i=1}^{N} k_i!} \cdot \prod_{i=1}^{N} p_i^{k_i}$$

where for the $N$ classes, the $N$ probabilities sum to unity, and the $N$ numbers of classification selections $k_i$ sums to the $n$ total selections:

$$\sum_{i=1}^{N} p_i = 1 \quad , \quad \sum_{i=1}^{N} k_i = n$$

As before, when the number of voters $n$ rises, it becomes more likely that the largest number of choices among the entries in $k = (k_i)$ will be for the category with the largest probability entry in $p = (p_i)$. If as before, we know the number of votes for each category as $k = (k_i)$, but we do not know the set of category probabilities $p = (p_i)$, then we may compute the distribution of those unknown probabilities by employing the Dirichlet distribution, which is the multinomial analogue of the Beta distribution. Here $f(p,\alpha)$ is the density of a probability vector $p = (p_i)$ given a vector of parameters $\alpha = (\alpha_i)$ where by convention, $\alpha_i = k_i + 1$.

$$f_{dirichlet}(\boldsymbol{p}; \boldsymbol{\alpha}) = \frac{((\sum_{i=1}^{N} \alpha_i) - 1)!}{\prod_{i=1}^{N}(\alpha_i - 1)!} \cdot \prod_{i=1}^{N} p_i^{\alpha_i - 1}$$

We proceed by computing the distribution of the probability vectors $p = (p_i)$, which is a multivariate distribution, and determine for each category $i$ the fraction of densities where the probability $p_i$ exceeds the probabilities of each of the other categories individually, thus establishing the most likely "true" category for the object being classified.

Previously in the two-class case, we declared a category to be most likely "true" if its probability exceeded that of the other category. Here, we may declare a category to be "true" if its probability exceeds that of all other categories combined. If a category has a majority vote of the entire population – that is, probability of greater than 50% (with enough votes counted to overcome our predetermined level of required certainty) – then it is selected the winner. We call this a *population preponderance*. This allows us to use what is called the marginal distribution of the density for each category, where the marginal distribution for a particular category is the distribution of that category summed over all possible values that may be assigned to any other category.

The marginal distribution of the Dirichlet distribution is the familiar Beta distribution. While it may seem intuitive to use the Beta distribution immediately to classify an object into one of *more* than two categories, a rigorous examination requires a formal statement that the Beta distribution is the marginal distribution of the Dirichlet distribution. This is proven by repeatedly leveraging what is known as the aggregation property of the Dirichlet distribution, which states that when any two categories in the distribution are combined into a single category, the resulting distribution is also Dirichlet, where the probability parameter $\alpha_i$ for the new combined category is simply the sum of the parameters for the replaced categories. This is repeated until only two categories remain, whereupon the result is identical to the Beta distribution. (The details of the proof of this description are omitted here, but abundant in the literature.)

In summary, for classification via voting, we accumulate votes until the cumulative Beta distribution for some category indicates that the category has a probability greater than 50% (a population preponderance) with a prescribed level of confidence. We note that the ability to do this is predicated upon the ability of our classifier to correctly identify samples from a particular radio at some fraction of the time that is greater than 50%

If the classifier cannot label the each category correctly more than 50% of the time, then a formulation of the classifier system in which the correct radio is identified at least *more often than any other* radio (though perhaps not 50% of the time) can be constructed, simply by ensuring that the Beta distributions are compared among all alternative radios individually, where the probabilities of error are then added (a conservative approach since errors can be concurrent rather than independent). While successful, this "population favored" method was not necessary for the present effort, because our classifier was able to obtain greater than 50% accuracy for all radio subsamples. The "population favored" method was, however, tested, and gave results in all cases identical to the population preponderance method, the test results for which are described below.

## 5 Bispectra of Signal Samples

Having discussed a voting scheme, we now define the voters. Earlier, we discussed that unintentional modulation is a ubiquitous characteristic of an emitter, and described the calculation of a bispectrum on a signal. Since we view the unintentional modulation of an emitter as ubiquitous throughout its signal, we may extract a random sample of a signal and expect that the bispectrum of the sample to depict the unintentional modulation of the emitter. Our paradigm for this discussion is to extract multiple random samples from a signal, compute the bispectrum of each sample, and deploy a trained deep learning system to create a vote for the emitter identity from that sample. Using the statistical techniques described above, we continue extracting samples and accumulating votes until an emitter identity is determined to any arbitrary prescribed degree of accuracy, which is the thesis of this paper.

One might question whether a high certainty that an emitter would be selected by a preponderance of a large population of samples, constitutes certainty about the identity of the emitter. The answer is yes. So far, we have written of "voters" as if their votes were matters of opinion, but at this point that analogy fails. What we have described as votes are in fact the bispectra of samples taken from the emitter whose identity is being determined, and those bispectra are evaluated by an algorithm trained specifically to identify emitters based on a preponderance of bispectra appearing to be characteristic of that emitter and not of other emitters. When the bispectra of all possible samples from an emitter are considered, having a preponderance identified with that emitter is precisely how we define certain identification, because we have trained and tested our classifier to uniquely identify each of our finite set of emitters on that basis.

## 6 Applying Computer Vision

We seek to use a state-of-the-art deep-learning computer vision system which has a low computational burden while maintaining sufficient accuracy to assign samples to emitters with reasonable certainty. For this we selected the EfficientNet [11] deep-learning system created by Google Labs, and found that the simplest version of this system, denoted EfficientNetB0, was sufficient for our purposes when coupled with a voting scheme as described above.

To construct our EfficientB0 system, we deleted the final 1000-category classification layer of a pre-trained EfficientNetB0 network and replaced it with two consecutive 256-node fully connected layers with rectified linear unit (ReLU) activation functions and L2=0.05 regularization, each preceded by a 50% dropout to improve generalizability, and appended a final 352-node softmax classification layer, where 352 = (16 radios) * (11 distances between radio and receiver) * (2 runs per setup) was the number of test cases in the published radio signal dataset. (ReLU, regularization, and dropout are techniques in neural network design which are outside the scope of this paper but described abundantly in the literature.)

EfficientNetB0 expects input dimensions of 224x224 for each input image, so that became the dimensions of our bispectral input; however, a bispectrum of a signal sample having only 224 sample points was not intuitively expected to have sufficient length to demonstrate an adequate duration of unintentional modulation, so we selected sample lengths of 1120 points, constructed the 1120x1120 bispectra, and downsampled to 224x224 bispectra using the summed power (amplitude squared) density within each 5x5 cell as the value for that cell. We rescaled the bispectral power quantities to integer values from 0 to 255, and then converted the resulting 2-D intensity plot to a 3-color plot by biasing the red, green, and blue color channels to high, medium, and low power levels respectively (using the Matlab "jet" colormap for reproduceability), thus constructing a 224x224x3 full color bispectrum image for each evaluated sample. While the details of these hyperparameters were chosen intuitively, and could have been chosen differently, subsequent testing revealed that adjusting them would not have significantly improved the results because arbitrary accuracy had already been obtained with minimal computational effort. The conversion from simple intensity (grayscale) to full color did, however, improve the subsequent results appreciably, since we could now leverage effectively the three color channels of the EfficientNetB0 convolutional neural network.

To create a bispectra data set, we extracted 200 samples from each of the two the recorded signals of 11 different SNR levels for each of the 16 radios, for a total of 704,000 samples, and repeated the process independently for validation and test data sets of 70,400 samples each. The samples constituted approximately 1.1 per cent of the available data for each signal. We then trained EfficientNetB0 using a categorical crossentropy optimizer (for a description of which the reader is referred to the machine learning literature) for approximately 200 epochs until the overall validation accuracy reached approximately 82%. (Higher accuracy was not necessary because the voting scheme described earlier does not require it. We presume that higher accuracies would be obtained with longer samples, larger bispectrum images, and so forth, but our objective was not high accuracy at this stage – rather, computational efficiency. The high accuracy is created later via the voting scheme.) We pause to note that we did test "whitening" the data set by applying PCA whitening, but this did not improve our overall results and was abandoned.

We note that the EfficientNetB0 deep learning network has 16,126,480 trainable parameters at single precision (32 bits per quantity) summing to about 62 megabytes, yet the RF signal data set consists of about 104 gigabytes of data, or over 1,700 times the size of learnable parameters in the network, so it seems unlikely that the network is essentially tabulating ("memorizing") the signal data in order to classify emitters. At the level of bispectra, our training data set consisted of 704,000 x 224 x 224 numbers prior to channelization, equating to about 35 billion numbers, or about 2,190 times the number of the trainable parameters in EfficientB0, leading to an analogous conclusion that the neural net could not have "memorized" the training data.

## 7 System Architecture

A graphical depiction of the computational steps in the system architecture is shown below.

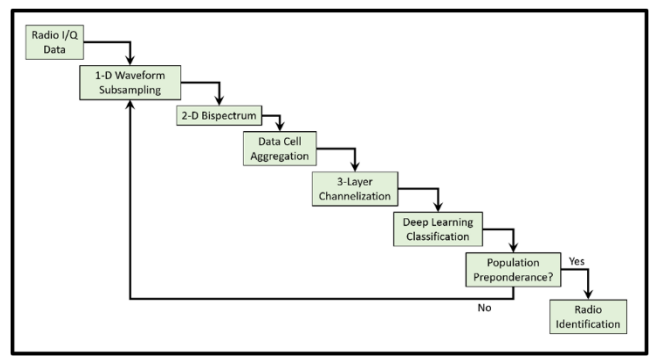

## 8 Testing

To identify a specific emitter, we extracted a random sample of its signal, computed the bispectrum, submitted the bispectrum to the deep-learning system for classification, and continued the process of extracting samples and classifying them until the cumulative Beta distribution function for any category indicated a population preponderance with a probability in excess of the preset accuracy requirement.graphical depiction of the computational steps in the system architecture

## 9 ACCURACY

To assess accuracy, we ran the model 19 times over each of the 352 test cases at 500 different Acceptable Error Thresholds ranging from 0.001 to 0.5, and obtained the Observed Error Rates for classification shown in the chart below for the 1.76 million total cases. A line of equality between the Observed Error Rate and the Specified Acceptable Error Threshold on this log-log plot is shown in red. Observed Error Rates do not appear on this plot for most Specified Acceptable Error Thresholds below 0.01 because in our simulations, the Observed Error Rates were zero, which is off the bottom of the chart.

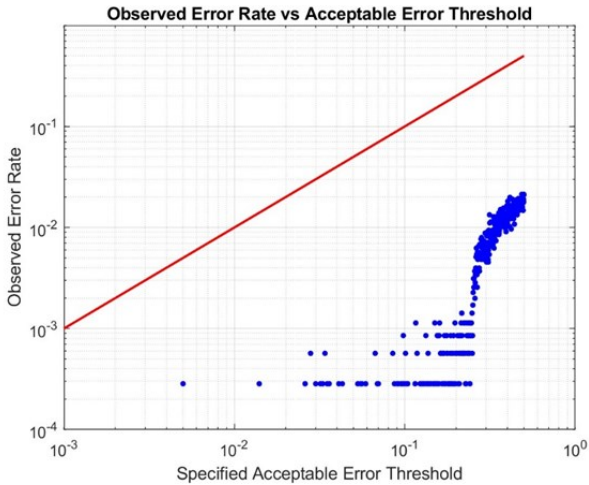

We observe that in typical cases, the Observed Error Rate is one or more orders of magnitude more favorable than the Acceptable Error Threshold. This is because the model continues to draw samples until its certainty meets or exceeds acceptability, but since the sampling process is inherently a discrete process, the certainty that is achieved will generally be more favorable than the acceptable limit, simply because it results from a discrete improvement at each step (e.g., in ascending though the integers until achieving an integer that is greater than pi, one must attain at least 4, significantly exceeding the threshold of pi simply because one must take a discrete step from 3 to 4 in order to meet the goal of being greater than or equal to the threshold of pi.)

## 10 CERTAINTY

We selected a variety of Acceptable Error Thresholds from $10^{-1}$ through $10^{-15}$ (the latter being a classification error rate of one part in one quadrillion) into 5000 logarithmically distributed values (to create a uniform distribution on a semilog plot), and for each Acceptable Error Threshold, we computed the radio identities to the prescribed level of accuracy for each of the 16 radios, 11 distances from the receiver, and 2 duplicate runs, for a total of 352 cases per Acceptable Error Threshold. In all of these simulation runs (352*5000) we did not observe any instances where the system failed to identify the correct radio, even though the samples drawn to make identifications were selected randomly from the test dataset (for each case). For each value of the threshold, we determined the largest number of samples that were required to reach the requisite level of certainty, across all radios at all distances from the receiver and at both of the two repeated runs. the prescribed accuracy rates. The maximum number of samples that needed to be drawn, as compared to the requisite Acceptable Error Thresholds, are shown below.

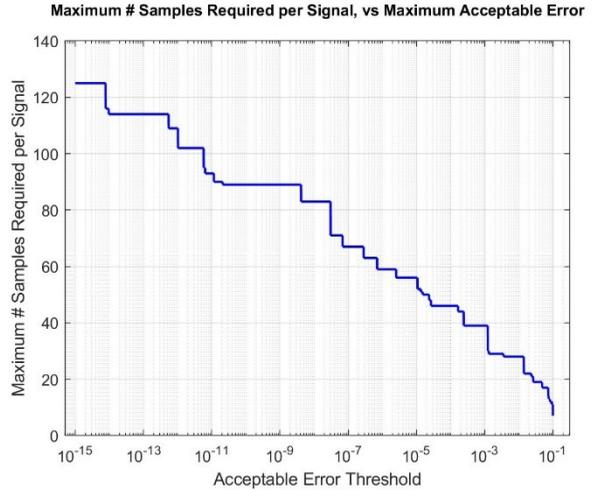

A linear least-squares best fit to the maximum number of samples required (NSAMP), as a function of the logarithm of the Acceptable Error Threshold (EAT), yields:

$$N_{SAMP} = -7.77 \cdot \log_{10} E_{AT} + 12.98$$

with a regression coefficient of causation R2 = 0.982. From this we see that a linear increase in the number of samples taken from a signal yields a logarithmic reduction in the error identifying the signal. In our application, *every additional eight samples taken from a signal yields a greater than 10-fold reduction in error* identifying that signal. This indicates that not only are very high levels of accuracy attainable, but they can be obtained by minimal increases in computational effort. Also, because each signal sample lasts for only 56 parts per million (ppm) of the entire signal duration, in the aggregate only a very small portion of each signal is actually needed to obtain the desired arbitrary level of accuracy. For example, from the chart above, in order to obtain an error rate of one in a trillion, it is only necessary to obtain about 110 samples in the worst case, or just over one half of one percent (0.0062) of the signal duration.

## 11 CONTRAST TO BAGGING AND BOOSTING

Techniques called "bagging" and "boosting" in the machine learning literature [12] refer to the use of several weak models together, to vote on identical items in order to obtain a smaller amount of error identifying each item. This is similar to averaging measurements, each with Gaussian standard error σ, to obtain an averaged estimate that will have standard error [13] of σ/√n . The technique described here is to use several samples together with a single model to obtain arbitrarily small levels of error on the order of $10^{(-n/8)}$. For example, "bagging" 72 models, each with Gaussian error, would be expected to reduce error by a factor of $1/\sqrt{72} = 0.118$, but using 72 samples in the presently described system will reduce error by a factor of $10^{(-72/8)} = 10^{(-9)} = 0.000000001$, so that in this example, the presently described model is over a hundred million times as effective in reducing error when compared against "bagging" or "boosting."

## 12 AN ANALOGY

Admittedly, these are "strong" results, but they arise because the bispectra detect unintentional modulations that are both unique to the various radios and also ubiquitous throughout their signals. By analogy, imagine that the Mona Lisa is copied by 16 talented fraudulent painters. To the naked eye, each copy looks exactly like the original. However, each painter actually has slightly different paint strokes, brush pressure, and slightly different colorings in the paint mix that the naked eye cannot detect. A single microscopist learns the difference painting tendencies in tiny swatches of each painting. The microscopist is then given several one-centimeter-square swatches from one of the fake paintings. Each swatch might not individually reveal which forgery it came from, but on average the swatches generally do tend toward a particular signature of the fraudster. The microscopist continues to vote on swatch after swatch until it becomes clear, to any desired level of certainty, which fraudster was the creator.

## 13 CONCLUSION

We have introduced a method of arbitrarily accurate classification using a system of voting upon subsamples with its attendant statistical interpretation, and applied it to specific emitter identification on a published dataset of physically recorded over-the-air signals from 16 ostensibly identical high-performance radios. For each signal subsample consisting of only about 56 parts per million of the original signal duration, we computed a two-dimensional full-color bispectrum from the one-dimensional I/Q data. We trained a three-channel computer vision deep learning convolutional neural network upon the images that resulted from 704,000 samples. We created a testing set of 70,400 unrelated samples and in testing, we performed voting on the test subsamples until any arbitrarily threshold level of accuracy in emitter identifications was obtained. This required minimal computation time because it was only necessary to evaluate on the order of about 20 to 50 testing subsamples to obtain error levels near zero. Previous efforts on the published dataset have resulted in error rates of approximately 1.4%, whereas this effort has resulted in an error rate arbitrarily close to zero, which appears to constitute a new state of the art.

**Michael C. Kleder** is a physicist and systems engineer with over 25 years of professional experience and patents in many countries, whose algorithmic software has been downloaded hundreds of thousands of times by researchers at national labs and private corporations. Mr. Kleder has an undergraduate degree in Business Management Science, and Master of Science degrees in both Physics and Systems Engineering, all from Virginia Polytechnic Institute and State University, where he is currently a doctoral candidate in the Industrial and Systems Engineering Department researching the application of systems thinking to hard problems in artificial intelligence
.